\documentclass[superscriptaddress,reprint]{revtex4-1}
\usepackage{amstext,amssymb}
\usepackage{graphicx}
\usepackage{xcolor}
\usepackage{amsmath}
\usepackage{amsfonts}

\newcommand\ep{\varepsilon}
\newcommand\tep{\varepsilon_0}
\newcommand\ps{s}
\begin{document}
\title{Boosting  thermoelectric efficiency using time-dependent control}
\author{Hangbo Zhou}
\affiliation{Department of Physics, National University of Singapore, Republic of Singapore 117551}
\affiliation{NUS Graduate School for Integrative Sciences and Engineering, National University of Singapore, Republic of Singapore 117456}
\author{Juzar Thingna}
\email[]{corresponding author: juzar@smart.mit.edu}
\altaffiliation{Present address: Singapore-MIT Alliance for Research and Technology (SMART) Centre, Singapore 138602}
\affiliation{Institute of Physics, University of Augsburg, Universit\"atstra{\ss}e 1, D-86135 Augsburg, Germany}
\affiliation{Nanosystems Initiative Munich, Schellingstra{\ss}e 4, D-80799 M\"unchen, Germany}
\author{Peter H\"{a}nggi}
\affiliation{Department of Physics, National University of Singapore, Republic of Singapore 117551}
\affiliation{Institute of Physics, University of Augsburg, Universit\"atstra{\ss}e 1, D-86135 Augsburg, Germany}
\affiliation{Nanosystems Initiative Munich, Schellingstra{\ss}e 4, D-80799 M\"unchen, Germany}
\affiliation{Centre for Phononics and Thermal Energy Science, School of Physics Science and Engineering, Tongji University, 200092 Shanghai, China}
\author{Jian-Sheng Wang}
\affiliation{Department of Physics, National University of Singapore, Republic of Singapore 117551}
\author{Baowen Li}
\email[]{baowen.li@colorado.edu}
\altaffiliation{Present address: Department of Mechanical Engineering, University of Colorado Boulder, CO 80309-0427, USA}
\affiliation{Department of Physics, National University of Singapore, Republic of Singapore 117551}
\affiliation{NUS Graduate School for Integrative Sciences and Engineering, National University of Singapore, Republic of Singapore 117456}
\affiliation{Centre for Phononics and Thermal Energy Science, School of Physics Science and Engineering, Tongji University, 200092 Shanghai, China}
\affiliation{Centre for Advanced 2D Materials and Graphene Research Centre, National University of Singapore, 6 Science Drive 2, Singapore 117546}

\date{24 August, 2015}

\begin{abstract}
Thermoelectric efficiency is defined as the ratio of power delivered to the load of a device to the rate of heat flow from the source. Till date, it has been studied in presence of thermodynamic constraints set by the Onsager reciprocal relation and the second law of thermodynamics that severely bottleneck the thermoelectric efficiency. In this study, we propose a pathway to bypass these constraints using a time-dependent control and present a theoretical framework to study dynamic thermoelectric transport in the far from equilibrium regime. The presence of a control yields the sought after substantial efficiency enhancement and importantly a significant amount of power supplied by the control is utilised to convert the wasted-heat energy into useful-electric energy. Our findings are robust against nonlinear interactions and suggest that external time-dependent forcing, which can be incorporated with existing devices, provides a beneficial scheme to boost thermoelectric efficiency.
\end{abstract}
\maketitle

The on-going advances of nano-structure engineering have re-energized the search for high-efficiency thermoelectric devices \cite{Giazotto2006,Dubi2011,Takabatake2014}. Till date, almost all studies on thermoelectricity are focused on finding high efficiency materials guided by the near equilibrium thermodynamic quantities like the Seebeck coefficient and the thermoelectric figure of merit $ZT$. These passive searches have reached saturation and the thermoelectric efficiency achieved thus far is still insufficient from a practical standpoint\cite{Vining2009}. This is primarily because, in the near equilibrium regime, the thermoelectric efficiency is limited by various thermodynamic constraints, namely, the second law of thermodynamics which imposes an unavoidable entropy flow and the Onsager reciprocal relation that connects the Seebeck and Peltier effects. 

In order to achieve high thermoelectric efficiency an active approach to overcome these thermodynamic obstacles is the need of the hour. A possible mechanism overcoming these thermodynamic constraints is to apply a time-dependent forcing to drive the system far from equilibrium. Unlike bulk materials, many nano-systems, such as quantum dots \cite{Lu2003, Benyamini2014}, single-electron-transistors \cite{Steele2009a}, and molecular junctions \cite{Liang2002,Zhitenev2002,Kohler2004,Crepieux2011,Chi2012}, can strongly interact with an externally applied control force. These systems have been the subject of intense theoretical investigations to better understand the mechanisms underlying electron \cite{Strass05} and heat \cite{Rey07, Arrachea07} transport in presence of a time-dependent control. Several applications such as overall device efficiency \cite{Juergens13}, thermopower \cite{Crepieux2011}, thermal refrigeration \cite{Rey07}, electron pumping \cite{Stefanucci08}, and heat pumping \cite{Segal06, Segal09} have also been studied in these systems to figure out the role of an external control. Despite these advances the study of dynamic thermoelectric efficiency has been highly non-trivial due to the breaking of thermodynamic constraints and the non-unique definitions of the thermodynamic quantities such as the Seebeck coefficient and the figure of merit $ZT$.

In this study we propose a dynamic theory of thermoelectric efficiency to overcome the present thermodynamic limitations such as the Onsager reciprocal relation. The main idea being that a strong external time-dependent control breaks the time-translational invariance of the system and pushes it in the far from equilibrium regime. This in turn causes a breakdown of the celebrated Onsager symmetry relation that allows for the possibility to boost the resulting efficiency. The boost can be as large as four times the near equilibrium value and vitally a large fraction of the supplied input energy is constructively utilized to enhance the thermoelectric performance. Thus, our novel approach makes available an extra knob to engineer high thermoelectric efficiency in nano-devices.
 
\noindent \textbf{Results}\\
\noindent \textbf{Dynamic theory of thermoelectricity.}
Since in the far from equilibrium regime the Seebeck coefficient and the figure of merit $ZT$ are ill-defined we establish a thermoelectric formalism based on the underlying time-dependent currents. This objective can be achieved through the evaluation of the Onsager transport matrix which relates the electron or heat current with the temperature or chemical potential bias. In the conventional, near equilibrium, formalism the transport-matrix coefficients are autonomous and constrained by various thermodynamic laws. However, a time-dependent control pushes the system far from equilibrium and results in the transport coefficients depending on the entire history of the applied protocol. Moreover the presence of a time-dependent control force causes the charging and discharging of the nano-system resulting in a time-varying thermoelectric current (known as the displacement current)\cite{Feve2007, Crepieux2011}, which will vanish exactly in the near-equilibrium scenario. 

\begin{figure}
\includegraphics[width=\columnwidth]{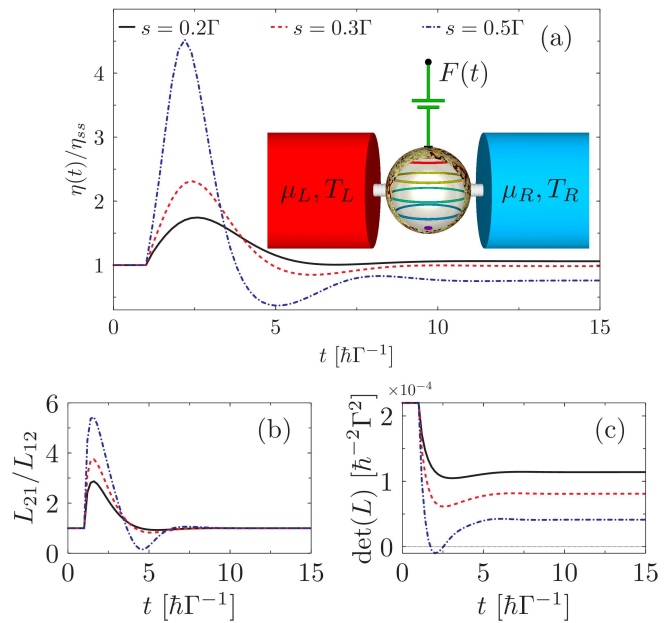}
\caption{\label{fig:figure1}\textbf{Efficiency and transport-matrix coefficients for non-interacting electrons.} (a): time evolution of the thermoelectric efficiency $\eta$ (normalized by steady-state efficiency $\eta(0)=\eta_{ss}$). Inset shows a typical set-up studied in this work of a multi-level system (depicted by coloured rings within a central sphere), acted upon by an external time-dependent control. (b): the entropy flow direction determined by $\mathrm{det}(L)$. (c): the Onsager reciprocal relation $L_{21}/L_{12}$. The control $F(t)=s \theta(t-1)$, $k_B(T_L + T_R)/2=0.1 \Gamma$, chemical potential $\mu_{L}=\mu_{R}=0$, and electron energy $\ep_0=0.5\Gamma$. The efficiency ratio is calculated with a bias $k_B\Delta T=0.02\Gamma$ and a load resistance $R_{\mathbb{L}}=15\hbar/e^2$.}
\end{figure}

In order to construe the above mechanism we consider a two-probe transport set-up consisting of a system connected to a left and right lead as depicted in the inset of Fig.~\ref{fig:figure1}(a). When the system is subjected to an external driving force $F(t)$ the left lead electron and heat current will be functions of the thermodynamic forces and the entire history of the applied protocol, $F(t')$, $t_{0} \leq t' \leq t$, with the starting time $t_0$ of the force protocol of otherwise {\it arbitrary} strength. In the linear response regime for the thermodynamic forces, namely, temperature difference $\Delta T/T$ and chemical-potential difference $\Delta\mu/e$ small, we Taylor-expand the currents at each instance of time as,
\begin{eqnarray}
I^L_e(t)&=&I^L_e(t)|^{\Delta T=0}_{\Delta\mu=0}+L_{11}[F]\frac{\Delta\mu}{e}+L_{12}[F]\frac{\Delta T}{T},\\
I^L_h(t)&=&I^L_h(t)|^{\Delta T=0}_{\Delta\mu=0}+L_{21}[F]\frac{\Delta\mu}{e}+L_{22}[F]\frac{\Delta T}{T}.
\end{eqnarray}
Above, the coefficients $L_{11}[F]=\frac{\partial I^L_e(t)}{\partial (\Delta \mu/e)}|_{\Delta T=0}$, $L_{12}[F]=\frac{\partial I^L_e(t)}{\partial (\Delta T/T)}|_{\Delta \mu=0}$, $L_{21}[F]=\frac{\partial I^L_h(t)}{\partial (\Delta \mu/e)}|_{\Delta T=0}$ and $L_{22}[F]=\frac{\partial I^L_h(t)}{\partial (\Delta T/T)}|_{\Delta \mu=0}$ indirectly depend on the entire history of the applied protocol via the currents $I^L_e(t)$ and $I^L_h(t)$. The zeroth order term $I^D_e(t)=I^L_{e(h)}(t)|^{\Delta T=0}_{\Delta\mu=0}$ arises only in the presence of an external control and is called the displacement current. Since the displacement current doesn't change sign under reversal of thermodynamic forces we can write the above Taylor expansion as a transport-matrix equation that reads,
\begin{equation}
\label{eq:Tmatrix}
\begin{pmatrix}I^\alpha_e(t) \\I^\alpha_h(t) \end{pmatrix} =\begin{pmatrix}L_{11}[F] & L_{12}[F] &\mathcal{L}^D_e[\cdot]\\L_{21}[F] & L_{22}[F] &\mathcal{L}^D_h[\cdot]\end{pmatrix}  \begin{pmatrix}\Delta^\alpha\mu/e \\\Delta^\alpha T/T \\F(t')\end{pmatrix}.
\end{equation}
The super-script $\alpha$ takes values $L$ and $R$ corresponding to the leads. The elementary charge $e > 0$ and $\Delta^L\mu=-\Delta^R\mu=\mu_L-\mu_R$ (similar interpretation for $\Delta^\alpha T$). Above $\mathcal{L}^{D}_{e(h)}[\cdot]$ represents displacement current kernel acting on the history of the applied protocol such that the displacement current $I^D_{e(h)}(t)=\mathcal{L}_{e(h)}^D[F]$.

For an undriven system; i.e. $F(t)=0 ~\forall t$, the displacement currents vanish, leaving only the biased currents, yielding a nonequilibrium steady state. Specifically, the transport matrix reduces to a commonly known, time-independent $2\times 2$ Onsager matrix $L=\begin{pmatrix}L_{11} & L_{12} \\L_{21} & L_{22} \end{pmatrix} $ and the transport coefficients are obeying the constraints of near equilibrium thermodynamic steady state transport; namely  the Onsager reciprocal relation are valid, imposing that $L_{21}/L_{12}=1$. Likewise, the second law of thermodynamics ensures a positive thermal conductance, or $\mathrm{det}(L)>0$ \cite{Takabatake2014}.

The primary result in this work is to obtain the transport coefficients under a time-dependent control. This can be achieved in the following manner: (i) We assume small thermodynamic forces for the temperature bias and the potential difference so that the relationship w.r.t to these forces stays linear.  (ii) The currents are evaluated (see below) at any time instant  $t$ as a function of the two small thermodynamic forces.  (iii) Then, setting $\Delta^{\alpha}\mu /e = 0$ the slope of the electron- (heat-) current w.r.t $\Delta^{\alpha}T/T$  yields $L_{12}[F]$ ($L_{22}[F]$) at the time  instant $t$. Likewise, for $\Delta^{\alpha}T/T=0$ we extract $L_{11}[F]$ and $L_{21}[F]$, respectively. The intercept of the electron (heat) current at time instant $t$ w.r.t $\Delta^{\alpha}T/T=0$ or $\Delta^{\alpha}\mu /e=0$ yields the contribution of the currents solely arising from the arbitrary driving $F(t)$; i.e. the displacement current.

In order to investigate the consequences of the time-dependent control on the thermoelectric efficiency we bias the system with a temperature difference $\Delta T$, connect a load of resistance $R_{\mathbb{L}}$ to the system and calculate the amount of power consumed by the load. We assume that the load is a pure resistor that cannot lead to charging effects due to the passage of electron current. Therefore, the amount of current passing through the load is  related to the bias and the transport matrix $L$. After accounting for the back-action from the load, the biased electron current reads $I_e(t)=L_{12}\Delta T/[T(1+M)]$, where $M = R_{{\mathbb L}}/R_{M}$ is the ratio of the resistances with $R_{M}\equiv L_{11}^{-1}$ being the resistance of the system. Hence the thermoelectric efficiency ratio of the heat-work conversion reads \cite{Goldsmid2009,Rowe2010},
\begin{equation}
\label{eq:efficiency}
\eta(t)=\frac{I_e^2R_{\mathbb{L}}}{\mathrm{det}(L)R_M\Delta T/T+L_{21}R_{M}I_e-I_e^2R_{M}/2}.
\end{equation}
Here, we have suppressed the explicit time-dependence in all terms on the r.h.s. for notational simplicity. The numerator $I_e^2R_{\mathbb{L}}$ is the useful power  on the load while the denominator is the heat extracted per unit time from the hotter lead. The extracted heat consists of three  contributions due to the entropy flow $\mathrm{det}(L)R_M\Delta T/T$, the Peltier heat due to the electron current $L_{21}R_{M}I_e$, and the Joule heating term $I_e^2R_{M}$ with the factor $-1/2$ indicating that half of the heat flows back to the hotter lead. In the nonequilibrium steady state this efficiency ratio will reduce to the standard formalism \cite{Dubi2011} where, $\mathrm{det}(L)R_M/T$ represents the thermal conductance, $L_{21}R_{M}$ is the Peltier coefficient, and the efficiency is directly related to the figure of merit $ZT$ provided that the Onsager reciprocal relation $L_{21}/L_{12}=1$ is satisfied. Since all the transport matrix coefficients [forming the numerator and denominator of Eq.~(\ref{eq:efficiency})] are affected by the time-dependent control, \emph{a priori} it is not clear if the control will have an overall enhancing or diminishing effect on the dynamic thermoelectric efficiency. Equation~(\ref{eq:efficiency}) represents one of our main results, generalizing the conventional thermoelectric theory to a dynamic one which can be applied to far-equilibrium non-steady-state regime. 

\noindent \textbf{Non-interacting electrons.}
As a proof of concept, we first consider a single electron quantum dot in the regime of strong Coulomb blockade. The time-dependent external force $F(t)$ causes charging and discharging on the system and the total Hamiltonian reads,
\begin{equation}
H=H_L+H_R+H_S(t)+H_T,
\end{equation}
where $H_{\alpha}=\sum_{k\in \alpha}\varepsilon_k c_k^\dagger c_k, \alpha=L,R$ is the Hamiltonian of the leads, $H_T=\sum_{\alpha=L,R}\sum_{k\in\alpha}V_k^{\alpha} c_k^\dagger d+\mathrm{h.c.}$ is the tunnelling Hamiltonian between the quantum dot and the leads, and the Hamiltonian of the quantum-dot system is
\begin{equation}
H_S(t)=\bigl[\ep_0+ F(t)\bigr]d^\dagger d\;.
\end{equation}
This quantum resonant model has been extensively studied in the context single-electron-transistors \cite{Stoof96, Brandes04, Steele2009a, Benyamini2014, Schaller14}, molecular junctions \cite{Park2000} or nano-wires \cite{Camalet2003, Kohler2005}. The energy level of the dot can be controlled  either via a time-dependent gate voltage \cite{Crepieux2011}, or via long-wavelength electromagnetic fields such as microwaves \cite{Platero04, Chi2012} or lasers \cite{Lehmann2003, Kohler2005}.

The result of heat-work conversion efficiency ratio of this model under step-like control is shown in Fig.~\ref{fig:figure1}. From Fig.~\ref{fig:figure1}(a) we detect large enhancements in the efficiency, upto a factor of 4 compared to the steady-state efficiency, as soon as the step-pulse is applied. After some relaxation time the values eventually saturate to the new steady state. Interestingly, the magnitude of $L_{21}/L_{12}$ [Fig.~\ref{fig:figure1}(b)] shows a profile similar to the efficiency ratio indicating that the breakdown of the Onsager reciprocal relations $L_{21}/L_{12} \neq 1$ and the efficiency enhancement are closely intertwined. Physically, when $L_{21}$ is not bounded by $L_{12}$, the contribution of the particle flow to the heat current can increase under the influence of  external driving. As a result the efficiency is boosted via increasing the useful heat (due to particle flow) while limiting the waste heat (due to entropy flow). To substantiate this claim we plot $\mathrm{det}(L)$ in Fig.~\ref{fig:figure1}(c). Because the $\mathrm{det}(L)$ is proportional to the entropy flow we  see that it decreases in the regime of efficiency enhancements. Importantly, for sufficiently strong driving ($s=0.5\Gamma$) we detect a regime with negative values for $\mathrm{det}(L)$, indicating a reversal of the entropy flow, even though the overall heat current still flows from the hot lead to the cold one.

\noindent \textbf{Harvested power.}
Supplementary to the colossal boosts in the heat-work conversion efficiency it is also crucial that most of the input power due to the control force is properly utilized to enhance the heat-work conversion efficiency. We analyse this using the harvested and input power
\begin{align}
\label{eq:powerinp}
\dot{w}_{\mathrm{harvest}}&=I_{e}^2(t)\, R_{\mathbb{L}}-I_{e}^2(t_{0})\, R_{\mathbb{L}}, \nonumber \\
\dot{w}_{\mathrm{input}}&=-2F(t)I_e^D(t)/e,
\end{align}
where $\dot{w}_{\mathrm{harvest}}$ is the power harvested from the enhancement of efficiency that is defined as the difference of the useful power consumed on the load under time-dependent control (first term) and the power from the steady-state contribution (second term). $\dot{w}_{\mathrm{input}}$ is the input power via the external control force, where the factor, $-F(t)/e$, represents the input voltage and $2I_{e}^{D}(t)$ is the total displacement current.

\begin{figure}
\includegraphics[width=\columnwidth]{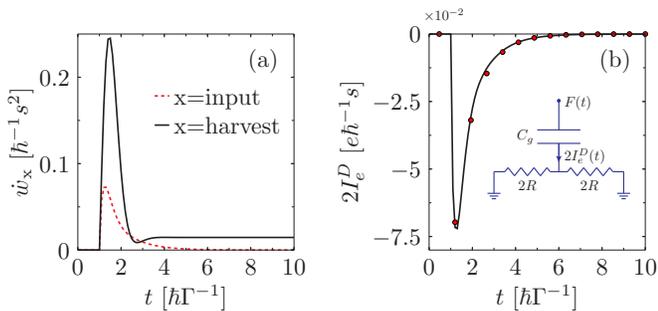}
\caption{\label{fig:figure2}\textbf{Power harvested, displacement current, and the intuitive resistance-capacitance model.} (a) the time-dependent harvested power (red dashed line) and the input power due to driving (black solid line). The system is a non-interacting electron model with load $R_{\mathbb{L}}=50\hbar/e^2$ and biased with $k_B(T_{L}+T_{R})/2= 1\Gamma$, $\Delta T=0.2\Gamma$, $\mu_{L}=\mu_{R}=0$. (b) displacement current (solid line) and the fit using the resistance capacitance model (red circles) for non-interacting electron model with $T_L=T_R=1\Gamma$, $\mu_L=\mu_R=0$, and $t_{0}=1\hbar/\Gamma$. The fitting parameters are $R=11.6\hbar/e^2$ and $\tau=1.06\hbar/\Gamma$. The common system parameters are: $s=0.001\Gamma$ and $\ep_0=2.5\Gamma$.}
\end{figure}

Figure~\ref{fig:figure2}(a) depicts that the harvested power can be much larger than the input power due to the control forcing. This feature occurs, even in the linear response regime, in the system-parameter regime when the steady-state efficiency is low, due to the low electron conductance, but the Seebeck coefficient itself remains large. Thus, the presence of time-dependent control constructively facilitates the movement of electrons and boosts the thermoelectric efficiency.

\noindent \textbf{Resistor-capacitor model.}
In order to better understand the displacement current in the high temperature and weak control limit, we propose an elementary resistor-capacitor model as illustrated in the inset of Fig.~\ref{fig:figure2}(b). The time-dependent control is acting on the gate with capacitance $C_g$ which can induce charging or discharging of the capacitor. This variation leads to a current generation which flows from the capacitance towards the leads which are represented as the two sink sources (the ground connection) in the circuit. The current generated solely depends on the time-dependent control and does not require a thermodynamic bias between the leads for its existence and is known as the displacement current $I^{D}_{e}(t)$. Due to dissipative effects the current experiences a total resistance $R$ while flowing from the capacitor to the leads.  

In case of the quantum dot model with non-interacting electrons subjected to a step-like gate control $F(t)=s\theta(t-t_0)$ the solution reads,
\begin{equation}
\label{eq:clasId}
I_e^D(t)=-\frac{s}{2eR}\theta(t-t_0)e^{-(t-t_0)/\tau}.
\end{equation} 
The intuitive picture for the displacement current above is based solely on circuit law considerations. Hence \emph{a priori} it is not clear if such a model is able to describe correctly a fully quantum mechanical system. In order to justify that this indeed is the case we use the parameters $R$ and $\tau$ from Eq.~(\ref{eq:clasId}) as variables and fit the equation to the fully quantum mechanical displacement current obtain via nonequilibrium Green's function. Figure~\ref{fig:figure2}(b) shows the NEGF calculation as a solid line and the fit via the red dots. The perfect fit gives us the parameters $\tau \approx\hbar/\Gamma$ which further strengthens our resistance-capacitance circuit model. This is because in an open dissipative quantum system one expects the relaxation time of the system to be inversely proportional to the sum of the coupling strengths of each lead $\Gamma^{-1}$ \cite{Weiss08}. Thus, the verification of our resistance-capacitance model gives an intuitive picture of the displacement current.

\noindent \textbf{Electron-phonon interaction.}
One of the main challenges for experimental devices to obtain enhanced efficiencies is the unavoidable presence of nonlinear interactions mainly arising due to the involvement of stray phonon modes\cite{Malen09}, which can drastically alter the thermoelectric efficiency. Hence, we consider a quantum dot interacting with a single phonon mode giving rise to the following electron-phonon interaction Hamiltonian,
\begin{equation}
H_S(t)=[\ep_0+F(t)]d^\dagger d+\omega_0 a^\dagger a +\lambda d^\dagger d(a^\dagger +a)\;.
\end{equation}
Here,  $a^\dagger$ and $a$ are creation and annihilation operators of the phonon, $\omega_0$ is the phonon angular frequency, $\lambda$ is the electron-phonon interaction strength and $F(t)$ represents the time-dependent control of the coherently driven quantum dot. The model manifests itself under various physical scenarios like in a nano-mechanical resonator \cite{Benyamini2014, Steele2009a,Zhou2015}, molecular junction \cite{Park2000, Lehmann2004}, and standard lattice vibration model \cite{Mahan2000}. Recently it was shown that a small amount of nonlinearity in this model can greatly suppress the steady-state efficiency \cite{Zhou2015}. Thus, the model serves as a perfect test bed to establish the robustness of our approach.

\begin{figure}
\includegraphics[width=\columnwidth]{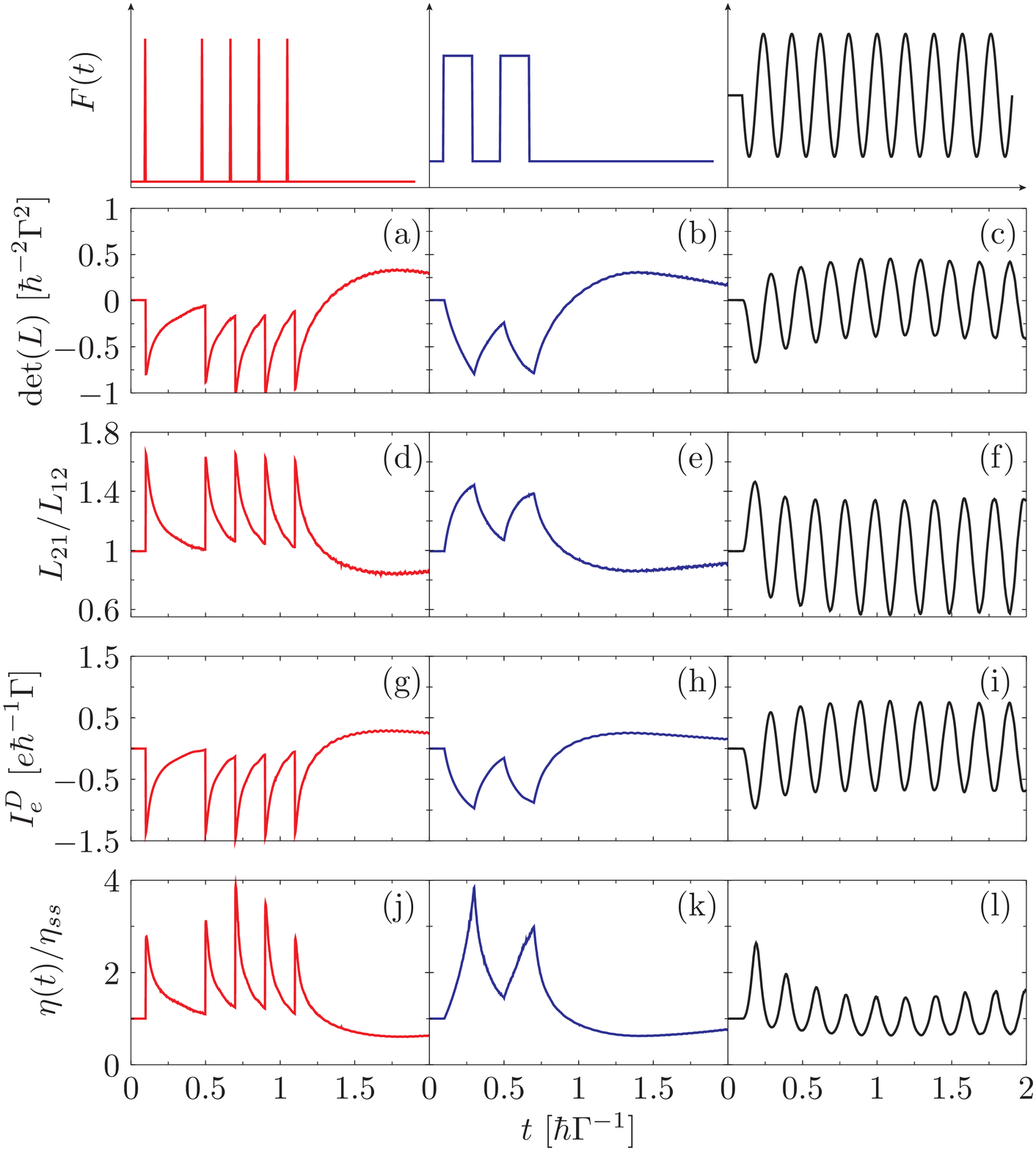}
\caption{\label{fig:figure3}\textbf{Efficiency and transport-matrix coefficients for electron-phonon interaction.} Plot of the entropy flow represented by $\mathrm{det}(L)$ (a, b, and c), the Onsager reciprocal relation $L_{21}/L_{12}$ (d, e, and f), the displacement current $I^{D}_e$ (g, h, and i), and the efficiency ratio normalized by the steady-state $\eta(t)/\eta_{ss}$ (j, k, and l) for the interacting electron model. The system is subjected to delta pulse driving $F(t)=s\sum_n\delta[\Omega (t- t_n)]$ with $\Omega t_n=\{1,5,7,9,11\}$ (a, d, g, and j), multi-step driving $F(\Omega t)=s$ when $\Omega t\in[1,3]\cup[5,7]$ and $F(t)=0$ elsewhere (b, e, h, and k) and a periodic sinusoidal drive $F(t)=2s\theta(t-t_{0})\sin(\Omega\pi t)$ (c, f, i, and l), where $\Omega=10\Gamma/\hbar$ and $t_{0}=0.1\hbar/\Gamma$. Other parameters are $k_B(T_{L}+T_{R})/2=1\Gamma$, $\Delta T=0.2\Gamma$, $\mu_{L}=\mu_{R}=0$, $\Gamma_{L}=\Gamma_{R}=\Gamma/2$, $\ep_0=2\Gamma$, $\omega_0=10\Gamma$, $\lambda=3\Gamma$, and $s=1\Gamma$.}
\end{figure}

In Fig.~\ref{fig:figure3} we depict the results for the interacting electron model. In case of the delta shape and square wave driving we modulate the system for sometime and then let it relax to reach its nonequilibrium steady state. Clearly the enhancement in the efficiency (as seen from the bottom row of Fig.~\ref{fig:figure3}) is observed even for a relative strong nonlinear interaction $\lambda$ as long as the system dynamics is time-dependent. The long-time limit, when the transient effects are wiped out, is easily recovered for the delta shaped and square wave driving giving $\eta(\infty)/\eta_{ss} =1$ [not depicted in Fig.~\ref{fig:figure3}]. Analogous to the non-interacting electron model, the enhancements are closely related to the breakdown of the Onsager reciprocal relation $L_{21}/L_{12}$ and the second law of thermodynamics $\mathrm{det}(L)$. Interestingly, external forcing alone is not sufficient to enhance the systems efficiency as seen from the case with a periodic sinusoidal driving where the efficiency even decreases when $L_{21}/L_{12}<1$. Thus, we speculate that although driving is a necessary condition to allow the breakdown of stringent constraints it does not suffice to enhance the efficiency of the device. One possible sufficient condition for an enhancement is the abrupt variation in the driving field which causes a sudden change of the charge occupation in the system. As a result the displacement current will be large (fourth row of Fig.~\ref{fig:figure3}).

\noindent \textbf{Discussion}\\
The performance of modern thermoelectric devices is measured in terms of its efficiency to convert waste heat energy into useful electric energy. Till date, most efforts towards efficiency enhancement are based on the search for suitable materials guided by the near-equilibrium thermodynamic constraints. These efforts have led to considerable improvements in the heat-work conversion efficiency but the quest for commercially feasible efficiency has been futile so far. 

In the present study, we provide an active and complimentary approach to the material search avenue. This is achieved by introducing a time-dependent control which pushes the system far from equilibrium and provides a rich playground without thermodynamic limitations. The control force can be fused with existing high efficiency devices and would allow us to further boost their efficiency by a factor of 4. The enhancements are robust and persist even in presence of nonlinear interactions indicating its usefulness in existing experimental set-ups.

Thus, our work opens up a whole new arena where we shift attention away from a material design perspective and focus on the non-trivial far from equilibrium regime which leads to smart device design. Overall the method presented herein provides a rigorous stepping stone which can have wide ranging impact for fields such as thermoelectric cooling \cite{Naik06, Zippilli09, Connell10, Santandrea11}, solar thermoelectric generation \cite{Goldsmith64, Kraemer11, Bell08}, and can be extended to the field of spin caloritronics to efficiently pump spin-currents using thermal gradients \cite{Bauer12}. 

\noindent \textbf{Methods}\\
\noindent \textbf{Nonequilibrium Green's function.}
For non-interacting electron in a quantum dot under step-like control with $F(t)=s\theta(t-t_0)$ within the wide-band approximation, i.e., $\Gamma_{\alpha}(\varepsilon)=\sum_{k\in\alpha}|V_k^{\alpha}|^2\delta(\varepsilon-\varepsilon_k)\equiv \Gamma/2$ ($\alpha = L,R$), an exact solution of the electron and heat currents can be obtained using the Landauer formalism via the NEGF approach \cite{Jauho1994, Crepieux2011}, reading
\begin{equation}
\label{eq:negf}
I^\alpha_{e(h)}(t)=-\sum_{\alpha'=L,R} \int_{-\infty}^\infty \frac{d\ep}{2\pi} f_{\alpha'}(\ep) K^{\alpha\alpha'}_{e(h)}(\ep,t),
\end{equation}
where the kernels
\begin{align}
K^{\alpha\alpha'}_{e}(\ep,t) &= e Z^{\alpha\alpha'}(\ep, t), \nonumber\\
K^{\alpha\alpha'}_{h}(\ep,t) &= \frac{\Gamma^{2}}{4} \mathrm{Im}\left\{A(\ep,t) \partial_t A^*(\ep,t)\right\} \nonumber \\
&+ (\ep-\mu_{\alpha'})Z^{\alpha\alpha'}(\ep, t), \nonumber\\
Z^{\alpha\alpha'}(\ep, t) &= \frac{\Gamma}{\hbar}\left[\delta_{\alpha,\alpha'}\mathrm{Im}\left\{A(\ep,t)\right\}+\frac{\Gamma}{4} |A(\ep,t)|^{2}\right].
\end{align}
The Fermi-Dirac distribution of the $\alpha$-th lead $f_\alpha(\ep)=[1+e^{\beta_\alpha(\ep-\mu_{\alpha})}]^{-1}$ with $\beta_\alpha=1/(k_BT_\alpha)$ and $A(\ep,t)$ is the spectral density,
 \begin{equation}
A(\ep,t)=\frac{\ep-\tep+i\Gamma/2-\ps e^{i(t-t_0)(\ep-\tep-\ps+i\Gamma/2)/\hbar}}{(\ep-\tep+i\Gamma/2)(\ep-\tep-\ps+i\Gamma/2)}
\end{equation}
Using Eq.~(\ref{eq:negf}) it is possible to obtain the displacement current, 
\begin{equation}
I^D_e=-\frac{\Gamma}{2\pi\hbar}\int_{-\infty}^\infty d\ep\Big\{\mathrm{Im}[A(\ep,t)]+\frac{\Gamma}{2}|A(\ep,t)|^2\Big\}f(\ep),
\end{equation}
by setting $f_L(\ep)=f_R(\ep)=f(\ep)$. For step-like driving we obtain the intuitive results where the displacement current vanishes in the long time limit. Subtracting the displacement currents from the total currents we obtain the biased currents,
\begin{eqnarray}
I^B_e(t)&=&\frac{\Gamma^2}{4\hbar}\int_{-\infty}^\infty \frac{d\ep}{2\pi} |A(\ep,t)|^2\big[f_L(\ep)-f_R(\ep)\big],\\
\label{eq:baisede}
I^B_h(t)&=&\frac{\Gamma}{4\hbar}\int_{-\infty}^\infty \frac{d\ep}{2\pi}\Big\{(\ep-\mu)\Gamma|A(\ep,t)|^2\nonumber\\
&+&\mathrm{Im}\left[A(\ep,t)\hbar\frac{\partial A^*(\ep,t)}{\partial t}\right]\Big\}\big[f_L(\ep)-f_R(\ep)\big].
\label{eq:biasedh}
\end{eqnarray}
The biased electron current $I^B_e(t)$ and the first term of the biased heat current $I^B_h(t)$ take the form of the Landauer formula which helps preserve the Onsager symmetry. The second contribution to the biased heat current arises only due to the presence of an explicit time-dependent, time-reversal breaking control that is responsible the breakdown of the Onsager symmetry.

Given the biased and displacement currents the NEGF formalism allows us to obtain a closed form expression of the efficiency $\eta(t)$ in the weak system-bath coupling and weak control limit. In order to achieve this goal we first simplify the efficiency given by Eq.~(\ref{eq:efficiency}) for maximum power output, i.e., $M = 1$ and negligible Joule heating (small $\Delta T$) to obtain
\begin{eqnarray}
\eta(t) = \frac{L_{12}^2}{4L_{11}L_{22}-2L_{12}L_{21}}\frac{\Delta T}{T}.
\label{eq:efficiency_mp}
\end{eqnarray}
The weak coupling limit (see Append. of Ref.~\cite{Juzar2012}) transforms the transmission $|A(\ep,t)|^2$ to a delta function and subsequently keeping only leading order terms in the control strength $s$ helps obtain the currents analytically in a closed form. These currents are then used to obtain the transport matrix coefficients $L_{11}, L_{12}, L_{21},$ and $L_{22}$ that result in the efficiency
\begin{eqnarray}
\eta(t) &=& \eta_{ss} + s \frac{\partial \eta_{ss}}{\partial \epsilon_0}\frac{f_2(t-t_0)}{f_1(t-t_0)}+s\tilde{\eta}(t-t_0)\frac{\Delta T}{T},
\label{eq:efficiency_t}
\end{eqnarray}
where $\eta_{ss}$ is the steady-state efficiency at time $t=0$ and $\tilde{\eta}(t-t_0)$ can be expressed in the high temperature limit as,
\begin{eqnarray}
\tilde{\eta}(t-t_0) &=& \frac{\pi}{2\hbar\left(\mu-\epsilon_{0}\right)}\frac{(t-t_0)e^{-\Gamma (t-t_0)/2}}{f_1(t-t_0)}.
\end{eqnarray}
The two functions are, $f_1(t-t_0) = e^{-\Gamma (t-t_0)} + f_2(t-t_0)$ and $f_2(t-t_0) = (1-e^{\Gamma (t-t_0)/2})^2$. Their ratio, $f_2(t-t_0)/f_1(t-t_0)$, monotonically increases from 0 to 1 as $(t-t_0)\rightarrow \infty$, thus providing no temporal boost in the efficiency arising from the second term in Eq.~(\ref{eq:efficiency_t}). The third contribution, i.e., $\propto \tilde{\eta}(t-t_0)$, arises due to the contribution that breaks the Onsager symmetry in the biased heat current [second term of Eq.~(\ref{eq:biasedh})] and is responsible for the temporal boost in the thermoelectric efficiency (in the leading order of $s$). This further strengthens our claim that it is indeed the breaking of Onsager symmetry leads to a boost in thermoelectric efficiency.

\noindent \textbf{Equations of motion for the resistor-capacitance model.}
We further elucidate on the resistor-capacitor model as shown in the inset of Fig.~\ref{fig:figure2}(b). Consider that the capacitor has a charge $Q$ then the voltage on its upper plate will be the sum of the voltage across the resistances and the voltage across the capacitor \cite{Horowitz15}, namely, 
\begin{equation}
-\frac{F(t)}{e} = \frac{Q(t)}{C_g}+2I^D_e(t) R.
\end{equation}
Above since $R$ is the total resistance, $2R$ will be the resistance of each resistor giving the voltage across each resistor as $2I^D_e(t) R$. Differentiating the above equation with respect to time we obtain
\begin{equation}
\label{eq:disI}
\dot{I}_e^D(t)+\frac{1}{\tau} I_e^D(t)+\frac{1}{2eR}\dot{F}(t)=0,
\end{equation}
where $\tau=RC_g$ represents the relaxation time of the leads. Above since the displacement current is due to the charging or discharging of the gate capacitance $C_g$ we have used $\dot{Q}(t)=2I^D_e(t)$ as the total displacement current. The solution to the differential equation reads
\begin{equation}
I_e^D(t)=\mathcal{L}^D_e[F]=-\frac{1}{2eR}\int_0^tdt' \dot{F}(t')e^{(t'-t)/\tau},
\end{equation}
The protocol $F(t)$ begins at $t_{0}$ ($0<t_{0}<t$) and ends at time $t$ and the displacement current depends on the complete history of the protocol.

\noindent \textbf{Quantum master equation.}
Due to the presence of nonlinear interactions in the system we resort to the time-dependent quantum master equation formulation to evaluate the currents. The formulation treats the nonlinear interactions exactly under an arbitrary forcing at the cost of a weak system-lead coupling. Following the standard scheme \cite{Breuer2007theory} the quantum master equation for the reduced density matrix $\rho(t)$ of the system reads
\begin{eqnarray}
\label{TDQME}
\frac{d\rho_{nm}}{dt}&=&-\frac{i}{\hbar}\Delta_{nm}(t)\rho_{nm}+\frac{1}{\hbar^2}\sum_{\substack{i,j \\ k,k'}}\mathcal{R}_{nmk'}^{ijk}\,\rho_{ij},
\end{eqnarray}
where the relaxation four-tensor
\begin{align}
\mathcal{R}_{nmk'}^{ijk} &=\left[Y^k_{ni}Y^{k'}_{jm}W^{kk'}_{ni}(t)-\delta_{j,m}\sum_{l} Y^{k}_{nl}Y^{k'}_{li}W^{kk'}_{li}(t)\right]+\mathrm{c.c.}.\nonumber
\end{align}
Above $\Delta_{ij}(t)=E_i(t)-E_j(t)$ is the energy spacing with $E_i(t)$ as the $i$-th instantaneous eigenenergy. Since the external time-dependent driving only modulates the eigenenergies, and does not affect the eigenstates of the system Hamiltonian, we use the eigenstates of the static Hamiltonian as our basis. The transition matrix $W_{ij}^{kk'}(t)=\int_{-\infty}^{t} dt' e^{-i\int_{t'}^t\Delta_{ij}(t'')dt''/\hbar}C^{kk'}(t-t')$, with the correlation function $C^{kk'}(t)= \protect\langle B^{k}(t)B^{k'}(0)\protect\rangle$. The vector-operators $Y$ and $B$ belong to the system and lead Hilbert space and appear in the tunnelling Hamiltonian; i.e., $Y=\{d,d^\dagger\}$ and $B=\{\sum_{\alpha=L,R}\sum_{k\in\alpha} V_k^{\alpha}c^\dagger_k,\sum_{\alpha=L,R}\sum_{k\in\alpha} V_k^{\alpha}c_k\}$ with $Y^{k}$ ($B^{k}$) denoting the $k$-th component of the $Y$ ($B$) vector. The operator $B(t)$ is the free-evolution of $B$ with the lead Hamiltonian $H_{L}+H_{R}$.

Generalizing the nonequilibrium steady-state formulation \cite{Juzar2012, Zhou2015, Thingna2014} to encompass time-dependent control $F(t)$  we obtain the expression for currents as,
\begin{equation}
\label{eq:qme}
I^L_{e(h)}(t) =\frac{2}{\hbar^2}\sum_{k,k'}\mbox{Im}\left\{\mbox{Tr}\left[\rho(t) Y^{k} Y^{k'}\mathcal{W}_{e(h)}^{kk'}(t)\right]\right\},
\end{equation}
where the electron or phonon hopping rates $\mathcal{W}_{e(h)}^{kk'}$ are defined, similar to the master equation, using the current-lead correlation functions $\mathcal{C}_{e(h)}^{kk'}(t)=\protect\langle B^{k}(t)\mathcal{B}_{e(h)}^{k'}(0)\protect\rangle$, where the operators $\mathcal{B}_e=\left\{\,e\sum_{k\in L}V_k^{L}c_k^\dagger,-e\sum_{k}V_k^{L}c_k\,\right\}$ and $\mathcal{B}_h=\left\{\,\sum_{k\in L}(\varepsilon_k-\mu_L) V_k^{L}c_k^\dagger,-\sum_{k}(\varepsilon_k-\mu_L) V_k^{L}c_k\,\right\}$. Once we know the time-dependent reduced density matrix of the system, we can deduce the displacement current as the time-derivative of the average charge on the dot as, $I_e^D(t)=-\dot{Q}/2$, where $Q=\mathrm{Tr}\big[\rho(t)d^\dagger d\big]$.

The above quantum master equation formalism is valid in the weak system-bath coupling limit and holds true for arbitrary control strength $s$ and electron-phonon interaction strength $\lambda$. Moreover, we do not resort  to the secular (or rotating wave) approximation and the master equation is kept semi-non-Markovian since time $t$ is explicitly present in the integral-limits of the transition matrix. The approach is robust to deal with nonlinear interactions exactly and thus allows to obtain the transport-matrix coefficients for strongly nonlinear systems.

~\\
\noindent \textbf{Acknowledgement}\\
J.-S.W. is supported by FRC grant R-144-000-343-112 and MOE grant R-144-000-349-112.\\
~\\
\noindent \textbf{Author contributions}\\
H.Z. derived the dynamic theory of thermoelectricity with assistance from J.T.. J.-S.W., P.H., J.T. and H.Z. derived the quantum master equation formalism. H.Z. carried out the numerical simulation. H.Z., J.T. and J.-S.W. analysed the results. P.H. was involved in several studies which led to this work and provided key contributions about the correct physical interpretation of the transport matrix due to time-dependent driving. B.L. proposed this project and along with J.-S.W. supervised the work at every stage. All authors contributed equally towards the discussion of the results and the presentation of the manuscript.\\
~\\
\noindent \textbf{Additional information}\\
\noindent \textbf{Competing financial interests:} The authors declare no competing financial interests.
\end{document}